\documentclass{entcs}
\usepackage{entcsmacro}
\def\lastname{Gacek, Miller, Nadathur}

\usepackage{proof}
\usepackage{xspace}
\usepackage{amssymb,amsmath}
\usepackage{hyperref}

\newcommand{\ignore}[1]{}

\newcommand{\logic}{${\cal G}$\xspace}
\newcommand{\hh}{$hH^2$\xspace}
\newcommand{\foldnb}{$FO\lambda^{\Delta\nabla}$\xspace}
\newcommand{\N}{{\rm I} \! {\rm N}}
\def\FOLDN{FO\lambda^{\Delta\N}}
\newcommand{\LG}{$LG^\omega$\xspace}

\newcommand{\lead}[1]{\smallskip\noindent{\bf #1}\quad}

\newcommand{\supp}{{\rm supp}}

\newcommand{\cut}{\hbox{\sl cut}}

\newcommand{\hsl}[1]{\hbox{\sl #1}}
\newcommand{\abs}[2]{\hsl {abs} \; #1 \; #2}
\newcommand{\app}[2]{\hsl {app} \; #1 \; #2}
\newcommand{\step}[2]{\hsl {step} \; #1 \; #2}
\newcommand{\steps}[2]{\hsl {steps} \; #1 \; #2}
\newcommand{\type}[1]{\hsl {type} \; #1}
\newcommand{\halts}[1]{\hsl {halts} \; #1}
\newcommand{\reduce}[2]{\hsl {reduce} \; #1\; #2}
\newcommand{\val}[1]{\hsl {value} \; #1}
\newcommand{\arr}[2]{\hsl {arr} \; #1 \; #2}
\newcommand{\of}[2]{\hsl {of} \; #1 \; #2}

\newcommand{\seq}[3]{\hsl {seq}_{#1} \; #2 \; #3}
\newcommand{\prog}[2]{\hsl {prog} \; #1 \; #2}
\newcommand{\member}[2]{\hsl {member} \; #1 \; #2}
\newcommand{\element}[3]{\hsl {element}_{#1} \; #2 \; #3}
\newcommand{\nat}[1]{\hsl {nat} \; #1}
\newcommand{\tuple}[1]{\langle #1 \rangle}
\newcommand{\cntx}[1]{\hsl {cntx} \; #1}
\newcommand{\subst}[3]{\hsl {subst} \; #1 \; #2 \; #3}

\newcommand{\name}[1]{\hsl {name} \; #1}
\newcommand{\fresh}[2]{\hsl {fresh} \; #1 \; #2}

\newcommand{\tlam}[3]{\lambda #1 \! : \! #2 . #3}

\newcommand{\defL}{\hsl {def}\mathcal{L}}
\newcommand{\defR}{\hsl {def}\mathcal{R}}

\newcommand{\tridot}{\!\Vdash\!}
\newcommand{\conc}[1]{(\,\Vdash\!\tuple{#1})}
\newcommand{\ctxconc}[2]{(#1\!\Vdash\!\tuple{#2})}

\newcommand{\ie}{{\em i.e.}}

\newcommand{\etc}{{\em etc}}

\begin{document}

\begin{frontmatter}
\title{Reasoning in Abella about Structural Operational Semantics
  Specifications}

\author{Andrew Gacek\thanksref{UMN}}
\author{Dale Miller\thanksref{LIX}}
\author{Gopalan Nadathur\thanksref{UMN}}

\thanks[UMN]{Department of Computer Science and Engineering,
  University of Minnesota, Minneapolis, MN 55455.}

\thanks[LIX]{INRIA Saclay - \^Ile-de-France \& LIX/\'Ecole
  polytechnique, Palaiseau, France}

\thanks[REALONES]{This work has been supported by 
INRIA through the ``Equipes Associ{\'e}es'' Slimmer, and by the NSF
Grant CCR-0429572 which includes funding for Slimmer. Opinions,
findings, and conclusions or  recommendations expressed in this papers
are those of the authors and do not necessarily reflect the views of
the National Science Foundation.}

\begin{abstract}
The approach to reasoning about structural operational semantics style
specifications supported by the Abella system is discussed. 
This approach uses $\lambda$-tree syntax to treat object language
binding and encodes binding related properties in generic judgments.
Further, object language specifications are embedded directly into the
reasoning framework through recursive definitions. The treatment of
binding via generic judgments implicitly enforces distinctness and
atomicity in the names used for bound variables. These properties
must, however, be made explicit in reasoning tasks.  This
objective can be achieved by allowing recursive definitions
to also specify generic properties of atomic predicates. The utility
of these various logical features in the Abella system is demonstrated
through actual reasoning tasks. Brief comparisons with a few other
logic based approaches are also made.


\end{abstract}
\end{frontmatter}

\section{Introduction}
\label{sec:intro}

This paper concerns reasoning about the
descriptions of systems that manipulate formal objects such as
programs and their specifications. 
A common approach to modelling the dynamic and
static semantics of these systems is to use a syntax-driven rule-based
presentation.   
These presentations can be naturally encoded as theories 
within a simple, intuitionistic logic.  
If the intuitionistic logic supports $\lambda$-terms and the
quantification of variables ranging over such terms, then it also
provides a convenient means for capturing binding notions in the
syntactic objects of interest; in particular, it facilitates the use
of the $\lambda$-tree approach to abstract syntax. 
A further benefit to using such a logic to encode semantic
specifications is that an immediate and effective animation of them is 
provided by logic programming systems such as $\lambda$Prolog
\cite{nadathur88iclp} and Twelf \cite{pfenning99cade}.

Given a logic-based specification of a formal system, establishing
properties of the system reduces to answering questions about what is
provable in the logic encoding the specification. Different
approaches can be adopted for this task. At one end, the
specification logic can be formalized and reasoned about within a
general purpose theorem-proving framework such as that provided by
Coq \cite{bertot04book} or Isabelle \cite
{nipkow02book}. At the other end, one can develop another logic,
often called a {\em meta-logic}, that is explicitly tuned to
reasoning about the specification logic. It is the latter approach
that we examine here. In particular, we expose its practical use
within the context of a specific theorem-proving system called Abella
\cite{gacek08ijcar}. 

The design of a logic that can act as a powerful and expressive 
  meta-logic has been the subject of much recent research
\cite{baelde07cade,gacek08lics,mcdowell02tocl,miller05tocl,tiu06lfmtp}.
The logics emanating from these studies share a common theme: they all
provide recursive definitions as a means for encoding specification
logics and some form of generic reasoning for modelling binding
notions at the meta level. We expose here an expressive and flexible
logic called \logic within this
framework. Abella is based on \logic but also provides special
support for the ways in which \logic is intended to be used in
meta-reasoning tasks. Our presentation pays attention to
the novel features of both \logic and Abella from this perspective.
Concreteness is provided by considering proofs of
evaluation, typing, and normalization properties of the $\lambda$-calculus. 

This paper is organized as follows.  The logic \logic is
summarized in Section~\ref{sec:logic} and its particular realization
in Abella is discussed in
Section~\ref{sec:arch}. Section~\ref{sec:example} illustrates the use
of Abella in a significant theorem-proving task, that of formalizing a
Tait-style proof of normalizability in the $\lambda$-calculus. 
Section~\ref{sec:future-work} points out limitations of
the currently implemented system.  Finally, in
Section~\ref{sec:related} we compare Abella-style reasoning with some
other approaches to the same kind of reasoning tasks.



\section{The Logical Foundation}
\label{sec:logic}

The logic \logic \cite{gacek08lics} which we use to formalize
arguments about structural operational semantics is based on an 
intuitionistic and predicative subset of Church's Simple Theory of
Types. Terms in \logic are monomorphically typed and are constructed
using abstraction and application from constants and (bound)
variables. The provability relation concerns terms of the
distinguished type $o$ that are also called 
formulas. Logic is introduced  by including special constants
representing the propositional connectives $\top$, $\bot$, $\land$,
$\lor$, $\supset$ and, for every type $\tau$ that does not contain
$o$, the constants $\forall_\tau$ and $\exists_\tau$ of type $(\tau
\rightarrow o) \rightarrow o$.  The binary propositional connectives
are written as usual in infix form and the expression $\forall_\tau
x. B$ ($\exists_\tau x. B$) abbreviates the formula $\forall_\tau
\lambda x.B$ (respectively, $\exists_\tau \lambda x.B$).  Type
subscripts are typically omitted from quantified formulas when their 
identities do not aid the discussion.

The standard treatment of the universal quantifier accords it an
extensional interpretation. When treating $\lambda$-tree syntax it is
often necessary to give importance to the form of the argument for a
statement like ``$B(x)$ holds for all $x$'' rather than focusing on
whether or not every instance of $B(x)$ is true. The $\nabla$ quantifier 
\cite{miller05tocl} is used to encode such generic
judgments. Specifically, we include the constants $\nabla_\tau$ of 
type $(\tau \rightarrow o) \rightarrow o$ for each type $\tau$ (not
containing $o$).  As with the other quantifiers, $\nabla_\tau x. B$
abbreviates $\nabla_\tau \lambda x. B$.

The \foldnb logic \cite{miller05tocl} incorporates $\nabla$
quantification into a sequent calculus presentation of intuitionistic
proof by attaching a local signature to every formula occurrence in a sequent.
We are interested here in considering also proofs that use
induction. In this situation, we are led naturally to including
certain structural rules pertaining to local 
signatures \cite{tiu06lfmtp}. Written at the level of formulas, 
these are the $\nabla${\em -exchange rule} $\nabla x\nabla y. F \equiv
\nabla y\nabla x. F$ and the $\nabla${\em -strengthening rule} $\nabla
x. F \equiv F$, provided $x$ is not free in $F$.  
If we adopt these rules, we can make all local signatures equal and 
hence representable by an (implicit) global binder. We shall refer to
these globally $\nabla$-bound variables as {\it nominal
  constants}.
Intuitively, one can think of nominal constants as denoting arbitrary,
unique names.
Notice that the exchange rule requires us to consider
atomic judgments as being identical if they differ by only
permutations of nominal constants.

The logic \logic uses the above treatment of the $\nabla$ quantifier
that was first introduced in the \LG system 
\cite{tiu06lfmtp}. Specifically, an infinite collection of nominal
constants are assumed for each type.  The set of all nominal
constants is denoted by $\mathcal{C}$. These constants are distinct
from the collection of usual, non-nominal constants denoted by
$\mathcal{K}$.  
We define the {\it support} of a term (or 
formula) $t$, written $\supp(t)$, as the set of nominal constants 
appearing in it. 
A permutation of nominal constants is a type preserving bijection
$\pi$ from $\mathcal{C}$ to $\mathcal{C}$ such that $\{ x\ |\ \pi(x)
\neq x\}$ is finite. Permutations are extended to
terms (and formulas), written $\pi . t$, as follows:
\begin{tabbing}
\qquad\=\kill
\> $\pi.a = \pi(a)$, if $a \in \mathcal{C}$ \qquad\qquad\=
$\pi.c = c$ if $c\notin \mathcal{C}$ is atomic \\
\> $\pi.(\lambda x.M) = \lambda x.(\pi.M)$
\> $\pi.(M\; N) = (\pi.M)\; (\pi.N)$
\end{tabbing}

\begin{figure*}[t]
\[
\infer[id_\pi]{\Sigma : \Gamma, B \vdash B'}{\pi.B = \pi'.B'}
  \quad
\infer[\cut]{\Sigma : \Gamma, \Delta \vdash C}
            {\Sigma : \Gamma \vdash B & \Sigma : B, \Delta \vdash C}
\]
\[
\begin{array}{cc}
\noalign{\smallskip}
\infer[\forall\mathcal{L}]{\Sigma : \Gamma, \forall_\tau x.B \vdash C}
      {\Sigma, \mathcal{K}, \mathcal{C} \vdash t : \tau &
       \Sigma : \Gamma, B[t/x] \vdash C} &
\infer[\forall\mathcal{R},h\notin\Sigma]
      {\Sigma : \Gamma \vdash \forall x.B}
      {\Sigma, h : \Gamma \vdash B[h\ \bar{c}/x]}
\\
\noalign{\smallskip}
\infer[\nabla\mathcal{L},a\notin \supp(B)]
      {\Sigma : \Gamma, \nabla x. B \vdash C}
      {\Sigma : \Gamma, B[a/x] \vdash C} &
\infer[\nabla\mathcal{R},a\notin \supp(B)]
      {\Sigma : \Gamma \vdash \nabla x.B}
      {\Sigma : \Gamma \vdash B[a/x]}
\\
\noalign{\smallskip}
\infer[\exists\mathcal{L},h\notin\Sigma]
      {\Sigma : \Gamma, \exists x. B \vdash C}
      {\Sigma, h : \Gamma, B[h\; \bar{c}/x] \vdash C} &
\infer[\exists\mathcal{R}]{\Sigma : \Gamma \vdash \exists_\tau x.B}
      {\Sigma, \mathcal{K}, \mathcal{C} \vdash t:\tau &
       \Sigma : \Gamma \vdash B[t/x]}
\end{array}
\]
\vspace{-0.25cm}
\caption{The core rules of \logic: the introduction rules for the
  propositional connectives are not displayed.} 
\label{fig:core-rules}
\end{figure*}

Figure \ref{fig:core-rules} presents a subset of the core rules for \logic;
the standard rules for the propositional connectives have been
omitted for brevity. 
Sequents in this logic have the form $\Sigma :
\Gamma \vdash C$ where $\Gamma$ is a set and the signature
$\Sigma$ contains all the free variables of $\Gamma$ and $C$.
In the rules, $\Gamma, F$ denotes $\Gamma \cup \{F\}$.
In the
$\nabla\mathcal{L}$ and $\nabla\mathcal{R}$ rules, $a$ denotes a
nominal constant of appropriate type. In the $\exists\mathcal{L}$
and $\forall\mathcal{R}$ rules, $\bar{c}$ is a listing of the
variables in $\supp(B)$ and $h\; \bar{c}$ represents the
application of $h$ to these constants; raising is used here
to encode the dependency of the quantified variable on
$\supp(B)$ \cite{miller92jsc}. The judgment $\Sigma, \mathcal{K},
\mathcal{C} \vdash t : \tau$ that appears in the  
$\forall\mathcal{L}$ and $\exists\mathcal{R}$ rules enforces the
requirement that the expression $t$ instantiating the quantifier in
the rule is a well-formed term of type $\tau$ constructed from the
variables in $\Sigma$ and the constants in ${\cal K} \cup {\cal C}$.  

Atomic judgments in \logic are defined recursively by 
a set of clauses of
the form $\forall \bar{x} .(\nabla \bar{z} . H) \triangleq B$: here
$H$ is an atomic formula all of whose free variables are contained in
either $\bar{x}$ or in $\bar{z}$ and $B$ is an arbitrary formula all
of whose free variables are also free in $\nabla \bar{z}. H$. The atom $H$ is
the {\em head} of such a clause and $B$ is 
its {\em body}. No nominal constant is permitted to appear in
either of these formulas. A clause of this form provides part of the
definition of a relation named by $H$ using $B$. The $\nabla$
quantifiers over $H$ may be instantiated by distinct nominal
constants.  The variables $\bar{x}$ that are bound by the
$\forall$ quantifiers may be instantiated by terms that depend on any
nominal constant except those chosen for the variables in $\bar{z}$.

Certain auxiliary notions are needed in formalizing the rules for
definitions in \logic. A {\it substitution} $\theta$ is a
type-preserving mapping from variables to terms such   
that the set $\{x\ |\ x\theta \neq x\}$, the {\em domain} of $\theta$, is
finite. 
\ignore{
Although
a substitution is extended to a mapping from terms to terms, formulas
to formulas, \etc, when we refer to its {\it domain} and {\it range},
we mean these sets for this most basic
function. 
}
A substitution is extended to a function from terms to terms
in the usual fashion and we write its application using a postfix notation.
If $\Gamma$ is a set 
of formulas then $\Gamma\theta$ is the set $\{J\theta\ |\ J \in
\Gamma\}$. If $\Sigma$ is a signature then $\Sigma\theta$ is the
signature that results from removing from $\Sigma$ the variables in
the domain of $\theta$ and adding the variables that are free in the
range of $\theta$.
Given a clause $\forall x_1,\ldots,x_n . (\nabla \bar{z} . H)
\triangleq B$, we define a version of it raised over the nominal
constants $\bar{a}$ and away from a signature $\Sigma$ as  
\begin{tabbing}
\qquad\=\kill
\>$\forall \bar{h} . (\nabla \bar{z} . H[h_1\; \bar{a}/x_1, \ldots,
h_n\; \bar{a}/x_n]) \triangleq B[h_1\; \bar{a}/x_1, \ldots, h_n\; \bar{a}/x_n],$
\end{tabbing}
where $h_1,\ldots,h_n$ are distinct variables of suitable type that do
not appear in $\Sigma$. 
Finally, given the sequent $\Sigma : \Gamma \vdash C$ and the nominal
constants $\bar{c}$ that do not appear in the support of $\Gamma$ or
$C$, let $\sigma$ be any substitution of the form
\begin{tabbing}
\qquad\=\kill
\>$\{h'\; \bar{c}/h\ |\ h \in \Sigma\ \mbox{and}\ h'\ \mbox{is a
variable of suitable type that is not in}\ \Sigma\}$.
\end{tabbing}
Then we call the sequent $\Sigma\sigma : \Gamma\sigma \vdash
C\sigma$ a version of $\Sigma : \Gamma \vdash C$ raised over
$\bar{c}$. 

\begin{figure}[t]
\[
\infer[\kern-1pt\defL]{\Sigma : A, \Gamma \vdash C}{\{\Sigma'\theta :
  (\pi.B')\theta, \Gamma'\theta \vdash C'\theta\}}
\qquad
\infer[\kern-1pt\defR]{\Sigma : \Gamma \vdash A}{\Sigma' : \Gamma'
  \vdash (\pi . B')\theta}
\]
\vspace{-0.5cm}
\caption{Rules for definitions}
\label{fig:def}
\end{figure}

The introduction rules for atomic judgments based on definitions are
presented in Figure~\ref{fig:def}. The $\defL$ rule has a set of
premises that is generated by considering each definitional clause of
the form $\forall \bar{x}.(\nabla \bar{z}. H) \triangleq B$ in the
following fashion. Let $\bar{c}$ be a list of distinct nominal
constants equal in length to $\bar{z}$ such that none of these
constants appear in the support of  $\Gamma$, $A$ or $C$ and let
$\Sigma' : A', \Gamma' \vdash C'$  denote a version of the lower
sequent raised over  $\bar{c}$. Further, let $H'$ and $B'$ be obtained
by taking the head and body of a version of the clause being
considered raised over $\bar{a} = \supp(A)$ and away from $\Sigma'$
and applying the substitution $[\bar{c}/\bar{z}]$ to them. Then the
set of premises arising from this clause are obtained by considering
all permutations $\pi$ of $\bar{a}\bar{c}$ and all substitutions
$\theta$ such that $(\pi. H')\theta = A'\theta$, with the proviso that
the range of $\theta$ may not contain any nominal constants.  
The $\defR$ rule, by contrast, has exactly one premise that is obtained
by using any one definitional clause. $B'$ and $H'$ are generated from
this clause as in the $\defL$ case, but $\pi$ is now taken to
be any one permutation of $\bar{a}\bar{c}$  and $\theta$ is taken
to be any one substitution such that $(\pi . H')\theta = A'$, again
with the proviso that the range of $\theta$ may not contain any
nominal constants. 

Some of the expressiveness arising from the quantificational structure
permitted in definitions in \logic is demonstrated by the following
definitional clauses: 
\begin{tabbing}
\qquad $(\nabla x. \name x) \triangleq \top$ \qquad\qquad
$\forall E. (\nabla x. \fresh x E) \triangleq \top$
\end{tabbing}
The $\nabla$ quantifier in the first clause ensures that 
{\sl name} holds only for nominal constants. Similarly, the relative
scopes of $\forall$ and $\nabla$ in the second clause
force {\sl fresh} to hold only between a nominal constant and a term
not containing that constant.

\clearpage

When \logic is used in applications, bound variables in syntactic
objects will be represented either explicitly, by term-level,
$\lambda$-bound variables, or implicitly, by nominal constants.
The equivariance principle for nominal constants realizes alpha
convertibility in the latter situation. Encoding bound variables by
$\lambda$-terms ensures that substitution is built-in and that
dependencies of subterms on bindings is controlled; specific
dependencies can be realized by using the device of raising. 
Definitions with $\nabla$ in the head allow for a similar control over
dependencies pertaining to nominal constants and raising can be used
to similar effect with these as well.

\ignore{
As it is stated, the set of premises in the $\defL$ rule arising from
any one definitional clause is potentially infinite because of the
need to consider every unifying substitution. It is possible
to restrict these substitutions instead to the members of a complete
set of unifiers. In the situations where there is a single most
general unifier, as is the case when we are dealing with
the higher-order pattern fragment \cite{miller91jlc}, the
number of premises arising from each definition clause is bounded by
the number of permutations. In practice, this number can be quite
small.
}

The consistency of \logic requires some kind of stratification
condition to govern the possible negative uses of predicates in the
body of definitions. There are several choices for such a condition. 
Rather than picking one in an {\it a priori} fashion, 
we will note relevant such conditions as needed. 


The final capability of interest is induction over
natural numbers. These numbers are encoded in \logic using the type
$nt$ and the constructors $z : nt$ and $s : nt \to nt$. Use of
induction is controlled by the distinguished predicate $\hbox{\sl nat}
: nt \to o$ which is treated by specific introduction rules. In
particular, the left introduction rule for {\sl nat} corresponds to
natural number induction.




\section{The Architecture of Abella}
\label{sec:arch}

Abella is an interactive theorem prover for the logic \logic. The
structure of Abella is influenced considerably by a two-level logic
approach to specifying and reasoning about computations. There is a
logic---the intuitionistic theory of second-order hereditary Harrop
formulas that we call \hh here---that provides a convenient vehicle for
formulating structural, rule-based characterizations of a variety of
properties such as evaluation and type assignment. An especially
useful feature of such encodings is that derivations within this
``specification'' logic reflect the structure of derivations in the
object logic.\footnote{Since \hh is a subset of $\lambda$Prolog
  \cite{nadathur88iclp}, it turns out that such specifications can
  also be compiled and executed effectively
  \cite{nadathur99cade}.} Now, the specification 
logic can be embedded into \logic 
through the medium of definitions. When used in this manner, \logic
plays the role of a reasoning or meta logic: formulas in \logic can
be used to encapsulate properties of derivations in the specification
logic and, hence, of computations in the object logic. By keeping the
correspondences simple, reasoning within \logic can be made to
directly reflect the structure of informal arguments relative
to the object logics. 

This two-level logic approach was enunciated by McDowell and Miller 
already in the context of the logic $\FOLDN$
\cite{mcdowell02tocl}. Abella realizes this idea using a
richer logic that is capable 
of conveniently encoding more properties of computations. As a
theorem prover, Abella also builds in particular properties arising
out of the encoding of the specification logic. We discuss these
aspects in more detail below.


\lead{The specification logic}
The formulas of \hh are given by the following mutually recursive
definitions:
\begin{tabbing} \qquad
$G \;=\; A \;|\; A \supset G \;|\; \forall_\tau x. G \;|\; G \land G$
\qquad\qquad
$D \;=\; A \;|\; G \supset D \;|\; \forall_\tau x. D$
\end{tabbing}
In these definitions, $A$ denotes an atomic formula and $\tau$ ranges
over types of order 0 or 1 not containing $o$. 
The sequents for which proofs are constructed in \hh are
restricted to the form $\Delta\longrightarrow G$ where  $\Delta$
is a set of $D$-formulas and $G$ is a $G$-formula.  
For such sequents, provability in intuitionistic logic is completely
characterized by the more restricted notion of (cut-free) uniform
proofs \cite{miller91apal}. In the case of \hh, every sequent
in a uniform proof of $\Delta\longrightarrow G$ is of the form
$\Delta,{\cal L}\longrightarrow G'$ for some $G$-formula $G'$ and for
some set of atoms $\cal L$.  Thus, during the search for a proof of
$\Delta\longrightarrow G$, the initial context $\Delta$ is {\em
  global}: changes occur only in the set of atoms on the left and
the goal formula on the right.

\begin{figure}
\[
\infer{\Gamma \vdash x : a}{x : a \in \Gamma} \hspace{.5cm}
\infer{\Gamma \vdash m\; n : b}{\Gamma
  \vdash m : (a \to b) & \Gamma \vdash n : a} \hspace{.5cm}
\infer[\mbox{$x$ not in $\Gamma$}]{\Gamma \vdash (\tlam
  x a r) : (a \to b)}{\Gamma, x : a \vdash r : b}
\]
\vspace{-0.75cm}
\caption{Rules for relating a $\lambda$-term to a simple type}
\label{fig:typing}
\begin{center}
\begin{tabular}{c}
$\forall m, n, a, b[\of m (\arr a b) \land
    \of n a \; \supset \; \of{(\app m n)} b]$\\
$\forall r, a, b[\forall x[\of x a  \supset 
    \of{(r \; x)}{b}] \supset \of{(\abs a r)}{(\arr a b)}]$
\end{tabular}
\end{center}
\vspace{-0.25cm}
\caption{Second-order hereditary Harrop formulas (\hh) encoding simply typing}
\label{fig:hhtyping}
\end{figure}

We briefly illustrate the ease with which type
assignment 
for the simply typed $\lambda$-calculus can be encoded in \hh.  There
are two classes of objects in this domain: 
types and terms. For types we will consider a single base type called
$i$ and the arrow constructor for forming function types.  Terms 
can be variables $x$, applications $(m\; n)$ where $m$ and $n$ are
terms,
and typed abstractions $(\tlam x a r)$ where $r$ is a term and $a$ is
the type of $x$. The standard rules for assigning types to terms are
given in 
Figure \ref{fig:typing}.  Object-level untyped $\lambda$-terms
and simple types can be encoded in a simply typed (meta-level)
$\lambda$-calculus as follows.  The simple types are built from the
two constructors $i$ and {\sl arr} and terms are built using the two
constructors {\sl app} and {\sl abs}. Here, the constructor {\sl abs}
takes two arguments: one for the type of the variable being abstracted
and the other for the actual abstraction.  
Terms in the specification logic contain binding and so there is no
need for an explicit constructor for variables.  Thus, the
(object-level) term $(\tlam f {i\to i} (\tlam x i (f\; x)))$ can be
encoded as the meta-level term $\abs {(\arr i i)} (\lambda f. \abs i
(\lambda x. \app f x))$.

Given this encoding of the untyped $\lambda$-calculus and simple types,
the inference rules of
Figure~\ref{fig:typing} can be specified by the \hh formulas in
Figure \ref{fig:hhtyping} involving the binary predicate {\sl of}.  Note
that this specification in \hh does not maintain an explicit context
for typing assumptions but uses hypothetical judgments instead.  Also,
the explicit side-condition in the rule for typing abstractions is not
needed since it is captured by the usual proof theory of the universal
quantifier in the \hh logic.


\lead{Encoding specification logic provability in \logic}
The definitional clauses in Figure~\ref{fig:seq} encode \hh
provability in \logic. In these and other such clauses in this paper,
we use the convention that capitalized variables are implicitly
universally quantified at the head. 
This encoding of \hh provability  derives from McDowell and Miller
\cite{mcdowell02tocl}.
As described
earlier,
uniform proofs in \hh contain sequents
of the form $\Delta,{\cal L}\longrightarrow G$ where $\Delta$ is a
fixed set of $D$-formulas and $\cal L$ is a varying set of atomic
formulas.  Our encoding uses the \logic predicate
{\sl prog} to represent the $D$-formulas in $\Delta$:
the $D$ formula $\forall\bar x.[G_1\supset\cdots\supset
G_n\supset A]$ is encoded as the clause $\forall \bar
x. \prog A (G_1\land\cdots\land G_n) \triangleq \top$ and $\forall
\bar{x}. A$ is encoded by the clause $\forall \bar
x. \prog A tt \triangleq \top$.
Sequents are encoded
using the atomic formula $(\seq N L G)$ where $L$ is a list encoding
the set of atomic formulas $\cal L$ and $G$ encodes the $G$-formula.
The argument $N$, written as a subscript, encodes the height of the
proof tree that is needed in inductive arguments.
The constructor $\langle \cdot \rangle$ is used to
inject the special type of atom into formulas.  To simplify notation,
we write $L \tridot G$ for $\exists n . \nat n \land \seq n L G$.
When $L$ is $nil$ we write simply $\,\tridot G$.

\begin{figure}[t]
\begin{tabbing}
\qquad\=\kill
\> $\element N B (B::L) \triangleq \top$ \qquad\=
$\element {(s\; N)} B (C::L) \triangleq \element N B L$\\[6pt]
\> $\member B L \triangleq \exists n . \nat n \land \element n B L$ \\[6pt]
\> $\seq N L \langle A \rangle \triangleq \member A L$ \\
\> $\seq {(s\; N)} L (B \land C) \triangleq \seq N L B \land \seq N L C$ \\
\> $\seq {(s\; N)} L (A \supset B) \triangleq \seq N {(A
  :: L)} B$ \\
\> $\seq {(s\; N)} L (\forall B) \triangleq \nabla x. \seq N L (B\; x)$ \\
\> $\seq {(s\; N)} L \langle A\rangle \triangleq \exists b. \prog A b
\land \seq N L b$\\
\> $\seq {(s\; N)} L \langle A\rangle \triangleq \prog A tt$
\end{tabbing}
\caption{Second-order hereditary Harrop logic in \logic}
\label{fig:seq}


\end{figure}

Proofs of universally quantified $G$ formulas in \hh are generic in
nature. A natural encoding of this (object-level) quantifier in the definition
of {\sl seq}  uses a (meta-level) $\nabla$-quantifier. 
In the case of proving an implication, the atomic assumption is
maintained in a list (the second argument of {\sl seq}).
The penultimate clause for {\sl seq} implements backchaining over a
fixed \hh  specification (stored as {\sl prog} atomic formulas).
The matching of atomic judgments to heads of
clauses is handled by the treatment of definitions in the logic
\logic, thus the penultimate rule for {\sl seq} simply performs this matching
and makes a recursive call on the corresponding clause body.

With this kind of an encoding, we can now formulate and prove in
\logic statements about what is or is not provable in \hh.
Induction over the height of derivations may be needed in such
arguments and this can be realized via natural number induction on $n$
in $\seq n L P$.  Furthermore, the $\defL$ rule encodes case analysis
in the derivation of an atomic goal, leading eventually to a
consideration of the different ways in which an atomic judgment may have been
inferred in the specification logic. 
Abella is designed to hide much of the details of how the {\sl seq}
and {\sl prog} specifications work and to reflect instead the
aggregate structure described here.

Since we have encoded the entire
specification logic, we can prove
general properties about it in \logic that can then be used in
reasoning about particular specifications.
In Abella, various such specification logic properties can be invoked
either automatically or through the use of tactics. For example, the
following property, which is provable in \logic, states the judgment
$\ell \tridot g$ is not affected by permuting, contracting, or
weakening the context of hypothetical assumptions $\ell$.
\begin{tabbing}\qquad
$\forall \ell_1, \ell_2, g. (\ell_1 \tridot g) \land (\forall e .
\member e \ell_1 \supset \member e \ell_2) \supset (\ell_2 \tridot g)$
\end{tabbing}
This property can be applied to any specification judgment
that uses hypothetical assumptions. Using it with the encoding of
typing judgments for the simply typed $\lambda$-calculus, for example,
we easily obtain that permuting, contracting, or weakening the typing
context of a typing judgment does not invalidate that judgment.

Two additional properties of our specification logic which are useful
and provable in \logic are called the {\em instantiation} and {\em cut}
properties. The instantiation property recovers the notion of
universal quantification from our representation of the specification
logic $\forall$ using $\nabla$. The exact property is
\begin{tabbing}\qquad
$\forall \ell, g. (\nabla x. (\ell\; x) \tridot (g\; x))
\supset \forall t. (\ell\; t) \tridot (g\; t).$
\end{tabbing}
Stated another way,
although $\nabla$ quantification cannot be replaced by $\forall$
quantification in general, it can be replaced in this way when dealing
with specification judgments. The cut property allows us to remove
hypothetical judgments using a proof of such judgments. This property
is stated as the formula
\begin{tabbing} \qquad
$\forall \ell_1, \ell_2, a, g. (\ell_1 \tridot \langle a\rangle) \land
(a :: \ell_2 \tridot g) \supset (\ell_1, \ell_2 \tridot g),$
\end{tabbing}
which can be proved in \logic: here,
$\ell_1, \ell_2$ denotes the appending of two contexts. As a concrete
example, we can again take our 
specification of simply typed $\lambda$-calculus and use the
instantiation and cut properties to establish a type substitution
property, \ie, if $\Gamma_1, x:a \vdash m : b$ and $\Gamma_2 \vdash n
: a$ then $\Gamma_1, \Gamma_2 \vdash m[x := n] : b$.


\lead{Encoding properties of specifications in definitions}
Definitions were used above to encode the specification logic and also
particular specifications in \logic. There is 
another role for definitions in Abella: they can be used also to
capture implicit properties of a specification that are needed in a
reasoning task. As an example, consider the encoding of type
assignment. Here, the instances of $(\seq N L G)$ that
arise all have $L$ bound to a list of entries of the form $(\of x t)$
where $x$ is a nominal constant that is, moreover, different from all
other such constants appearing in $L$. Observing these properties is
critical to 
proving the uniqueness of type assignment. Towards this end, we may
define a predicate {\sl cntx} via the following clauses: 
\begin{tabbing}
\qquad\=\kill
\>$\cntx nil \triangleq \top$\qquad\qquad $(\nabla x. \cntx (
(\of x T)::L)) \triangleq \cntx L$
\end{tabbing}
Reasoning within \logic, it can now be shown that $L$ in every $(\seq
N L G)$ atom whose proof is considered always satisfies the property
expressed by {\sl   cntx} and, further,
if $L$ satisfies such a property then the uniqueness of type
assignment is guaranteed.  


\lead{Induction on definitions}
The logic \logic supports induction only over natural numbers. Thus
the definitions of {\sl element} and {\sl seq} in Figure~\ref{fig:seq}
both make use of a natural number 
argument to provide a target for induction. In Abella, such
arguments are unnecessary since the system implicitly assigns such an
additional argument to all definitions.  Thus when we refer
to induction over a definition we mean induction on the implicit
natural number argument of that definition.



\section{Example: Normalizability in the Typed
  \texorpdfstring{$\lambda$-}{Lambda }Calculus}
\label{sec:example}



\begin{figure}[t]
\begin{tabbing}
\qquad\=\kill
\> $\forall a, r [\val (\abs a r)]$\\[6pt]
\> $\forall m, n, m' [ \step m m' \supset \step {(\app m n)} (\app {m'}
n) ]$\\
\> $\forall m, n, n' [ \val m \land \step n n' \supset \step {(\app m n)}
(\app m n') ]$\\
\> $\forall a, r, m [ \val m \supset \step {(\app {(\abs a r)} m)} (r\;
m)]$\\[6pt]
\> $\forall m [\steps m m]$ \qquad\qquad\=
$\forall m, n, p [\step m p \land \steps p n \supset \steps m n]$\\[6pt]
\> $\type i$
\> $\forall a, b [\type a \land \type b \supset \type (\arr a b)]$\\[6pt]
\> $\forall a, b, m, n[\of m (\arr a b) \land
\of n a \supset \of{(\app m n)} b]$\\
\> $\forall a, b, r[\type a \land \forall x[\of x a \supset \of{(r \;
  x)}{b}] \supset \of{(\abs a r)}{(\arr a b)}]$
\end{tabbing}
\vspace{-0.25cm}
\caption{Specification of simply-typed $\lambda$-calculus}
\label{fig:spec}
\end{figure}

In order to illustrate the strengths and weaknesses of Abella, we
detail in this section a proof of normalizability for the call-by-value,
simply typed $\lambda$-calculus (sometimes also called ``weak
normalizability'').  We follow here the proof presented in
\cite{pierce02book}.  Stronger results are possible for the full, 
simply typed $\lambda$-calculus, but the one at hand suffices
to expose the interesting reasoning techniques. The proof under
consideration is based on Tait's logical relations argument
\cite{tait67jsl} and makes use of simultaneous substitutions.

Figure~\ref{fig:spec} contains the specification of call-by-value
evaluation and of simple typing for the $\lambda$-calculus.
Values are recognized by the predicate {\sl
  value}. Small-step evaluation is defined by {\sl step}, and a
possibly zero length sequence of small steps is defined by {\sl
  steps}. The predicate {\sl type} recognizes well-formed types, and
{\sl of} defines the typing rules of the calculus. 
A noteworthy aspect of the specification of the {\sl of} predicate is
that it uses the {\sl type} predicate to ensure that types
mentioned in abstraction terms are well-formed: a fact used in later
arguments. 
\ignore{
This is necessary to
do because later arguments will depend on the precise shape of this
type component.
One deviation of
this specification from its standard presentation is that the {\sl of}
predicate refers to the {\sl type} predicate to ensure that types
mentioned in abstraction terms are well-formed. This is required since
the object-logic is viewed as untyped from the perspective of our
meta-logic, and since we will later require very precise information
about the objects being manipulated.
}

The goal of this section is to prove weak normalizability, which we can
now state formally in our meta-logic as follows:
\begin{tabbing} \qquad
$\forall M, A. \conc{\of M A} \supset \exists V . \conc{\steps M V}
\land \conc{\val V}.$
\end{tabbing}
The rest of this section describes definitions and lemmas necessary to
prove this formula. In general, almost all results in this section
have simple proofs based on induction, case analysis, applying lemmas,
and building results from hypotheses. For such proofs, we will omit the
details except to note the inductive argument and key lemmas used. The
full details of this development are available in the software distribution of
Abella.


\lead{Evaluation and typing}
Definitions can be used in Abella to introduce useful intervening
concepts. One 
such concept is that of halting. We say that a term $M$ halts if it
evaluates to a value in finitely many steps and we define a predicate
capturing this notion as follows:
\begin{tabbing} \qquad
$\halts M \triangleq \exists V . \conc {\steps M V} \land \conc{\val V}.$
\end{tabbing}

An most important property about halting is that it is
invariant under evaluation steps (both forwards and backwards). Using
the abbreviation $F \equiv G$ for $(F \supset G) \land (G \supset F)$,
we can state this property formally as
\begin{tabbing} \qquad
$\forall M, N. \conc{\step M N} \supset (\halts M \equiv \halts N).$
\end{tabbing}
This result is immediate in the backward direction, \ie, $\halts N
\supset \halts M$. In the forward direction it requires showing that
one step of evaluation is deterministic:
\begin{tabbing} \qquad
$\forall M, N, P. \conc{\step M N} \land \conc{\step M P} \supset N = P.$
\end{tabbing}
This formula is proved by induction on the height of the derivation of
either one of the judgments involving the {\sl step} predicate.

A standard result in the $\lambda$-calculus, which we will need later, is
that one step of evaluation preserves typing. This is stated
formally as
\begin{tabbing} \qquad
$\forall M, N, A. \conc{\step M N} \land \conc{\of M A} \supset
\conc{\of N A}.$
\end{tabbing}
The proof of this formula uses induction on the height of the
  derivation of the judgment involving the {\sl step} predicate. An
  interesting case in this proof is when $\step M 
N$ is $\step {(\app {(\abs B R)} P)} (R\; P)$ for some $B$, $R$, and
$P$, \ie, when $\beta$-reduction is performed. Deconstructing the
typing judgment
\begin{tabbing} \qquad
$\conc{\of {(\app {(\abs B R)} P)} A}$
\end{tabbing}
we can deduce
that $\conc{\of P B} $ and $\ctxconc{(\of x B) :: nil}{\of {(R\; x)} A}$
where $x$ is a nominal constant. Here we use the instantiation
property of our specification logic to replace $x$ with $P$ yielding
$\ctxconc{(\of P B) :: nil}{\of {(R\; P)} A}$. Next we apply the cut
property of our specification logic to deduce $\conc{\of {(R\; P)} A}$
which is our goal.

Finally, we note that the contexts which are constructed during the
proof of a typing judgment always have the form $(\of {x_1} {a_1}) ::
\ldots :: (\of {x_n} {a_n}) :: nil$
where the $x_i$'s are distinct nominal constants
and the $a_i$'s are valid types.  We introduce the following formal
definition of {\sl cntx} to exactly describe such contexts:
\begin{tabbing} \qquad
$\cntx nil \triangleq \top$ \qquad\qquad
$(\nabla x. \cntx ((\of x A) :: L)) \triangleq \conc{\type A} \land \cntx L$
\end{tabbing}
Note, $\nabla$ in the definition head ensures that the
$x_i$'s are distinct nominal constants.


\lead{The logical relation}
The difficulty with proving weak normalizability directly is that the
halting property is not closed under application, \ie, $\halts M$ and
$\halts N$ does not imply $\halts (\app M N)$. Instead, we must
strengthen the halting property to one which includes a notion of
closure under application. We define the logical relation {\sl reduce}
by induction over the type of a term as follows:
\begin{tabbing}
\qquad\=$\reduce M (\arr A B)$ \= \kill
\> $\reduce M i $ \> $\triangleq \conc{\of M i} \land \halts M$ \\
\> $\reduce M (\arr A B)$ \> $\triangleq$ \=
$\conc{\of M (\arr A B)} \land \halts M \land {}$ \\
\>\>\> $\forall N. (\reduce N A \supset \reduce {(\app M N)} B)$
\end{tabbing}
Note that {\sl reduce} is defined with a negative use of itself. Such
a usage is permitted in \logic only if there is a stratification
condition that ensures that there are no logical cycles in 
the definition. In this case, the condition to use is obvious: the
second argument to {\sl reduce} decreases in size in the recursive use.

Like {\sl halts}, the {\sl reduce} relation is preserved by evaluation:
\begin{tabbing} \qquad
$\forall M, N, A. \conc{\step M N} \land \conc{\of M A} \supset
(\reduce M A \equiv \reduce N A).$
\end{tabbing}
This formula is proved by induction on the definition of {\sl reduce},
using the lemmas that {\sl halts} is preserved by evaluation and {\sl
  of}\/ is preserved by evaluation.

Clearly {\sl reduce} is closed under application and it implies the
halting property, thus we strengthen our desired weak
normalizability result to the following:
\begin{tabbing} \qquad
$\forall M, A. \conc{\of M A} \supset \reduce M A.$
\end{tabbing}
In order to prove this formula we will have to induct on the height of
the proof of 
the judgment $\conc{\of M A}$. However, when we consider the case that
$M$ is an abstraction, we will not be able to use the inductive
hypothesis on the body of $M$ since {\sl reduce} is defined only on
closed terms, \ie, those typeable in the empty context. The standard
way to deal with this issue is to generalize the desired formula to say
that if $M$, a possibly open term, has type $A$ then each closed
instantiation for all the free variables in $M$, say $N$, satisfies
$\reduce N A$. This requires a formal description of simultaneous
substitutions that can ``close'' a term.


\lead{Arbitrary cascading substitutions and freshness}
Given $\ctxconc{L}{\of M A}$, \ie, an open term and its typing context, we
define a process of substituting each free variable in $M$ with a
value $V$ which satisfies the logical relation for the appropriate
type. We define this {\sl subst} relation as follows:
\begin{tabbing}
\qquad\=$(\nabla x.$ \= $\subst {((\of x A) :: L)} {(R\; x)} M)$ \=\kill
\> \> $\subst {nil} M M \triangleq \top$ \\
\> $(\nabla x.$ \> $\subst {((\of x A) :: L)} {(R\; x)} M) \triangleq$ \\
\> \hspace{3cm} $\exists V .~ \reduce V A \land \conc{\val V} \land
\subst L {(R\; V)} M$
\end{tabbing}
By employing $\nabla$ in the head of the second clause, we are able to
use the notion of substitution in the meta-logic to directly and
succinctly encode substitution in the object language. Also note
that we are, in 
fact, defining a process of cascading substitutions rather than
simultaneous substitutions. Since the substitutions we define (using
closed terms) do not affect each other, these two notions of
substitution are equivalent. We will have to prove some part of this
formally, of course, which in turn requires proving results about
the (non)occurrences of nominal constants in our judgments. The results in
this section are often assumed in informal proofs.

One consequence of defining cascading substitutions via the notion of
substitution in the meta-logic is that we do not get to specify where
substitutions are applied in a term. In particular, given an
abstraction $\abs A R$ we cannot preclude the possibility
that a substitution for a nominal constant in this term will affect
the type $A$. 
Instead, we must show that well-formed types cannot contain free
variables which can be formalized as
$\forall A. \nabla x. \conc{\type (A\; x)} \supset \exists B.~ A =
\lambda y. B$.
This formula essentially states that any well-formed type which
possibly depends on a nominal constant $x$ 
must depend on it only in a vacuous way.

The above result about types assumes that judgments concerning {\sl
  type} occur in an empty context. Now, such judgments actually enter
  the picture through uses of the specification logic rule for {\sl
  of}\/ that deals with the case of abstractions.
This means that we have to consider judgments involving {\sl type} that
  have a context meant to be used 
in judgments involving the {\sl of} predicate. To use the result we
  have just established, we must show that these 
  contexts can be 
  ignored. We formalize this as
$\forall L, A.~\cntx L \land \ctxconc{L}{\type A} \supset \conc{\type A}$,
a formula that can be proved using induction on the proof of the judgment
$\ctxconc{L}{\type A}$. In the base case we must establish 
$\forall L, A.~\cntx L \land \member {(\type A)} L \supset \bot$,
which is proved by induction on the proof of {\sl member}.

Another necessary result is that in any provable judgment of the form
$\ctxconc{L}{\of M A}$, any nominal constant (denoting a free
variable) in $M$ must also occur in 
$L$, \ie,
\begin{tabbing} \qquad
$\forall L, R, A. \nabla x.~ \cntx L \land \ctxconc{L}{\of {(R\; x)} (A\; x)}
\supset \exists M.~ R = \lambda y. M$
\end{tabbing}
The proof is by induction on the height of the derivation of the
judgment involving {\sl of}. In the base case, we need that
an element of a list cannot contain any nominal constant which does
not occur in the list, \ie, 
$\forall L, E. \nabla x. ~\member {(E\; x)} L \supset \exists F.~ E =
\lambda y. F$.
This formula is proved by induction on {\sl member}.

We next show that typing judgments produce well-formed types by proving
\begin{tabbing} \qquad
$\forall L, M, A.~ \cntx L \land \ctxconc{L}{\of M A} \supset \conc{\type A}.$
\end{tabbing}
The induction here is on the height of the derivation of the judgment
involving {\sl of} and the base case is
$\forall L, M, A.~ \cntx L \land \member {(\of M A)} L \supset
\conc{\type A}$,
which is proved by a simple induction on {\sl member}.

Given our repertoire of results about the occurrences of nominal
constants in judgments, we can now prove fundamental properties of
arbitrary cascading substitutions.  The first property 
states that closed terms, those typeable in the empty context, 
are not affected by substitutions, \ie,
\begin{tabbing}\qquad
$\forall L, M, N, A.~ \conc{\of M A} \land \subst L M N \supset M = N.$
\end{tabbing}
The proof here is by induction on {\sl subst} which corresponds to
induction on the length of the list $L$. The key step within the proof
is using the lemma that any nominal constant in the judgment
$\conc{\of M A}$ must also be contained in the context of that
judgment. Since the context is empty in this case, there are no
nominal constants in $M$ and thus the substitutions from $L$ do not
affect it.

We must show that our cascading substitutions act compositionally on
terms in the object $\lambda$-calculus. This is stated formally for
application as follows:
\begin{tabbing}
\qquad
$\forall L, M, N, R . $ \=
$\cntx L \land \subst L {(\app M N)} R \supset$ \\
\> $\exists M', N'.~ R = \app {M'} N' \land
\subst L M M' \land \subst L N N'$.
\end{tabbing}
This is proved by induction on {\sl cntx}, which amounts to induction
on the length of the list $L$. For abstractions we prove the
following, also by induction on {\sl cntx}:
\begin{tabbing}
\qquad
$\forall L, M, R, A .~
\cntx L \land \subst L {(\abs A M)} R \land \conc{\type A} \supset$ \\
\qquad\qquad\qquad
$\exists M'.~ R = \abs A M' \land (\forall V.$ \= $\reduce V A \land
\conc{\val V} \supset$ \\
\> $\nabla x.~ \subst {((\of x A) :: L)} {(M\; x)} (M'\; V))$.
\end{tabbing}
Here we have the additional hypothesis of $\conc{\type A}$ 
to ensure that the substitutions created from $L$ do not affect $A$.
At one point in this proof we have to show that the order in which
cascading substitutions are applied is irrelevant. The key to showing
this is realizing that all substitutions are for closed terms. Since
closed terms cannot contain any nominal constants, substitutions do
not affect each other.

Finally, we must show that cascading substitutions 
preserve typing. Moreover, after applying a full cascading substitution
for all the free variables in a term, that term should now be typeable
in the empty context:
\begin{tabbing} \qquad
$\forall L, M, N, A.~\cntx L \land \subst L M N \land \ctxconc{L}{\of M A}
\supset \conc{\of N A}.$
\end{tabbing}
This formula is proved by induction on {\sl cntx} and 
by using the
instantiation and cut properties of our specification logic.


\lead{The final result}
Using cascading substitutions we can now formalize the generalization
of weak normalizability that we described earlier: given a (possibly open)
well-typed term, every closed instantiation for it satisfies the
logical relation {\sl reduce}\/:
\begin{tabbing} \qquad
$\forall L, M, N, A .~ \cntx L \land \ctxconc{L}{\of M A} \land \subst L M N
\supset \reduce N A.$
\end{tabbing}
The proof of this formula is by induction on the height of the
derivation of the typing
judgment $\ctxconc{L}{\of M A}$. The inductive cases are fairly
straightforward using the compositional properties of cascading
substitutions and various results about invariance under evaluation.
In the base case, we must prove
\begin{tabbing} \qquad
$\forall L, M, N, A.~ \cntx L \land \member {(\of M A)} L \land \subst L
M N \supset \reduce N A,$
\end{tabbing}
which is done by induction on {\sl cntx}.
Weak normalizability is now a simple corollary where we take $L$ to be
$nil$. Thus we have proved 
$\forall M, A. \conc{\of M A} \supset \halts M$.



\section{Assessment and Future Work}
\label{sec:future-work}

The Abella system has been tested with several prototypical
examples; details are available with the system distribution. These
experiments indicate considerable promise for the two-level logic
based approach in reasoning about formal systems. However, the
experiments have also revealed some issues with Abella at a
practical level. We discuss these below and suggest work aimed at
addressing them.

\ignore{
We use this section to assess our style of reasoning within a
two-level logical system, specifically focusing on the current
weaknesses. We fold this into a corresponding discussion of future
work to overcome these issues.
}

\lead{Base case lemmas}
Every lemma whose proof uses induction on a specification
logic judgment with a non-empty context requires another lemma to be
proved for the base case where that judgment follows because
it is in the context. This creates mundane overhead. The work in these
base case lemmas consists of a simple induction over the length
of the context. Support for richer tactics for induction on
specification judgments  might lead to more user friendly behavior in
such cases. 

\lead{Types in specifications}
The specification logic is embedded as an untyped logic in
\logic. This is usually not an issue: specification logic judgments
themselves impose type restrictions on terms. For example, the typing
judgment $\of M A$ holds only if $M$ is a $\lambda$-term. However,
sometimes explicit type judgments---such as the judgment {\sl type}
for recognizing well-formed simple types---are required in specifications.
One possibility that is being considered for addressing the typing
issue that is of an implementation such as Abella automatically generating
recognizer predicates based on type information. These predicates
could then be implicitly attached to all declarations of meta-level
variables. 

\lead{Different specification logics}
Currently, Abella has built into it exactly one
specification language (\hh) and exactly one proof system for it
(uniform proofs).  Certain application areas might benefit from having
other proof systems for intuitionistic logic available as well as
other specification logics.  For example,
linear logic specification languages \cite{hodas94ic,miller96tcs} can be
used to provide declarative specifications of the operational
semantics of programming languages that contain features such as
references, exceptions, and concurrency. Thus, McDowell and Miller
\cite{mcdowell02tocl} presented a {\sl seq}-like predicate
for a subset of intuitionistic linear logic that they used to 
specify the operational semantics of a simple functional
language extended with references and to then prove a
subject-reduction theorem for that language.  It would be natural 
to consider extending the specification logic in Abella to be all of
intuitionistic linear logic (or, in fact, all of linear logic) since
this would enhance that logic's expressiveness a great deal. Such an
extension could be designed so that if a given specification did not
employ the novel linear logic connectives, then the encoding of {\em
  seq} would modularly revert back to that of intuitionistic logic.



\section{Related Work}\label{sec:related}



\lead{Nominal logic approach}
The Nominal package for Isabelle/HOL automates a process of defining
and proving standard results about $\alpha$-equivalence classes
\cite{urban05cade}. This allows for formal reasoning over objects with
binding which is close to informal reasoning.  One
drawback of the nominal approach is that it does not provide a notion
of substitution, and thus users must define their own substitution
function and prove various properties relating to it. A proof of weak
normalizability for the simply typed $\lambda$-calculus has been
conducted with the nominal package \cite{narboux08nominal},
and in this case a notion of simultaneous substitution is used. For
the nominal approach, this extended notion of substitution can be
defined directly since one works with $\alpha$-equivalence classes and
not higher-order terms as in our case. Additionally, the cost of
defining and reasoning about simultaneous substitution is not a
significant step up from what is already required for standard
substitution in the nominal approach.

The specification language for the nominal package is functions and
predicates over $\alpha$-equivalence classes. This language does not
have a built-in notion of hypothetical judgments which are typically
useful for describing structural rules over objects
with binding. For example, by encoding the simply typed
$\lambda$-calculus in our specification language using hypothetical
judgments for typing assumptions, we derive a type substitutivity
property as consequence of general instantiation and cut properties of
the logic, see Section
\ref{sec:arch}.
In the nominal approach, such a proof must be conducted manually.


\lead{Twelf}
The Twelf system \cite{pfenning99cade} uses LF terms and types for a
specification language \cite{harper93jacm} and the meta-logic ${\cal
  M}_2^{+}$ \cite{schurmann00phd} for reasoning. The primary
difference between the Twelf approach and ours is that the ${\cal
  M}_2^{+}$ meta-logic is relatively weak in expressive power. For
instance, it is restricted to $\Pi_2$ formulas (\ie, $\forall\exists$
formulas) and lacks logical connectives such as conjunction,
disjunction, and implication. Despite these restrictions, the
meta-logic is expressive enough for most common reasoning tasks and
has been very successful in practice. Another significant difference
is that ${\cal M}_2^{+}$ is designed with an inherent notion of a
global hypothetical context. Thus the meta-logic builds in some notion
of which judgments can depend on assumptions of other judgments. This
is less of a concern in our approach since each judgments has its own
local context.

Due to the $\Pi_2$ restriction of the meta-logic ${\cal M}_2^{+}$, it
is not possible to encode a direct proof of weak normalizability for the
simply typed $\lambda$-calculus using a logical relations argument.
Recently, however, an indirect proof was completed using an
intermediate {\em assertion logic} which has enough richness to encode
the proper logical relation \cite{schurmann08lics}. This is a
useful technique for extending the expressive power of the Twelf
system, but it comes with the cost of moving from a two-level logic
approach to a three-level logic approach.

\lead{Locally nameless}
The locally nameless representation for syntactic objects with binding
is a first-order approach using de Bruijn indices for bound
variables and names for free variables.
This balance between two representational techniques has been used
successfully in practice \cite{aydemir08popl}.
Our approach to representation can be seen as a meta-level version of
this balance where we use (meta-level) $\lambda$-terms to represent
explicitly bound variables and (meta-level) nominal constants for
implicitly bound variables (\ie, free variables).
With this understanding, the trade-off between the first-order and
meta-level approaches to bound/free variable representation is that the former
works with existing theorem provers while the latter has substitution
and equivariance built-in.





\bibliographystyle{alpha}
\bibliography{../references/master}

\end{document}